\documentclass[useAMS,usenatbib]{mn2e}
\usepackage{multirow} 
\usepackage{epsfig}

\title{The influence of the turbulent perturbation scale on prestellar core fragmentation and disk formation.}
\author[S.~Walch, A.~Whitworth, P.~Girichidis]
{S. Walch$^{1}$ \thanks{E-mail: Stefanie.Walch@astro.cf.ac.uk}, A. P. Whitworth$^{1}$, P. Girichidis$^{1,2}$\\
$^{1}$School of Physics \& Astronomy, Cardiff University, 5 The Parade, Cardiff CF24 3AA, Wales, UK\\
$^{2}$Institut f\"ur Theoretische Astrophysik, Albert-Ueberle-Str. 2, 69120 Heidelberg, Germany }

\begin{document}
\date{Accepted 2011 August 31; Received 2011 March 24; in original form }
\pagerange{\pageref{firstpage}--\pageref{lastpage}} \pubyear{2009}
\maketitle

\begin{abstract}
The collapse of weakly turbulent prestellar cores is a critical stage in the process of star formation. Being highly non-linear and stochastic, the outcome of collapse can only be explored theoretically by performing large ensembles of numerical simulations. Standard practice is to quantify the initial turbulent velocity field in a core in terms of the amount of turbulent energy (or some equivalent) and the exponent in the power spectrum ($n \equiv -d\log P_k/d\log k$). In this paper, we present a numerical study of the influence of the details of the turbulent velocity field on the collapse of an isolated, weakly turbulent, low-mass prestellar core. We show that, as long as $n\ga3$ (as is usually assumed), a more critical parameter than $n$ is the maximum wavelength in the turbulent velocity field, $\lambda_{_{\rm MAX}}$. This is because $\lambda_{_{\rm MAX}}$ carries most of the turbulent energy, and thereby influences both the amount {\em and} the spatial coherence of the angular momentum in the core. We show that the formation of dense filaments during collapse depends critically on $\lambda_{_{\rm MAX}}$, and we explain this finding using a force balance analysis. We also show that the core only has a high probability of fragmenting if $\lambda_{_{\rm MAX}} > R_{_{\rm CORE}}/2$ (where $R_{_{\rm CORE}}$ is the core radius); that the dominant mode of fragmentation involves the formation and break-up of filaments; and that, although small protostellar disks (with radius $R_{_{\rm DISK}}\la20\,{\rm AU}$) form routinely, more extended disks are rare. In turbulent, low-mass cores of the type we simulate here, the formation of large, fragmenting protostellar disks is suppressed by early fragmentation in the filaments. \\
\end{abstract}

\begin{keywords}
hydrodynamics -- stars: formation -- stars: circumstellar matter -- turbulence  -- infrared: stars.
\end{keywords}

\section{Introduction}

Although there is strong observational evidence that circumstellar disks are formed during the early (Class 0 and Class I) phases of protostellar evolution, estimates of their masses and extents are uncertain by up to a factor of 10 \citep{Jorgensen2009}. Since protostellar disks that are sufficiently massive and extended to fragment might be an important site for forming low-mass stars, brown dwarfs and planetary mass objects \citep[e.g.][]{Stamatellos2010}, it is important to understand the circumstances under which such disks can form.

In this paper, we study the formation of protostellar disks in collapsing, weakly turbulent cores, and evaluate the influence of the scale of turbulent perturbations and the net core angular momentum. \footnote{Note that we are here using the term {\it turbulence} in the loose sense of random (and here statistically isotropic) macroscopic motions over a range of length-scales, and not in the more restrictive sense of {\it fully developed turbulence}, in which energy is injected on large length-scales and cascades through a large inertial range of length-scales, before being dissipated on much smaller length-scales \citep[e.g.][]{Kritsuk2007}. Here, {\it turbulence} is simply a device for seeding a star-forming cloud or core with the fluctuations that will eventually lead to fragmentation; the term is routinely used in this loose sense by those who simulate collapse and fragmentation.} In contrast to models based on rigid rotation \citep{Walch2009, Machida2010, Machida2011}, it has never been demonstrated unambiguously that the net angular momentum in a {\it turbulent} core \citep{Dib2010} significantly affects the size of the protostellar disk that it spawns. Rather, previous numerical models \citep[e.g.][]{Goodwin2004a, Walch2010} find no correlation between core angular momentum and disk size, suggesting that disk size is determined by details of the specific turbulent velocity field. However, previous work \citep{Klessen2000, Fisher2004, Goodwin2004a, Goodwin2004b, Matzner2005, Kratter2006, Goodwin2006, Krumholz2007, Attwood2009, Walch2010} has not investigated fully the parameter space used to initialize turbulent prestellar cores.

A random, Gaussian, turbulent velocity field is characterized -- in a statistical sense -- by five parameters. (i) Some measure of the total amount of turbulent energy, for example the root-mean-square turbulent velocity, $v_{_{\rm RMS}}$, or the mean turbulent Mach number, $\bar{\cal M}$. Observations \citep{Goodman1998, Barranco1998} suggest that turbulent velocities in low-mass cores are approximately sonic ($\bar{\cal M}\sim 1$). (ii) The partition of energy between solenoidal and compressive modes \citep[e.g.][]{Federrath2008}, where the statistical equilibrium distribution is 2:1 solenoidal to compressive. (iii) The slope of the turbulent velocity power spectrum $n$, where $P_k \propto k^{-n}$, and $n$ is typically between $3$ and $4$  \citep[][note that here $n$ is defined so that -- for example -- the Kolmogorov scaling index corresponds to $n\!=\!11/3$]{BurkertBodenheimer00}. (iv) The wavelength of the largest turbulent perturbation, $\lambda_{_{\rm MAX}}\equiv 2R_{_{\rm CORE}}/k_{_{\rm MIN}}$. (v) The wavelength of the smallest turbulent perturbation, $\lambda_{_{\rm MIN}} \equiv 2R_{_{\rm CORE}}/k_{_{\rm MAX}}$. Previous studies of low-mass core collapse -- and of star cluster formation from the collapse of larger, more turbulent molecular cloud cores, e.g. \citet{Bate2003, Bonnell2003, Bate2009a, Bate2009b} -- have not explicitly specified the last two parameters $(\lambda_{_{\rm MAX}},\lambda_{_{\rm MIN}})$. In this paper we evaluate their influence in more detail.

As long as (a) the power spectrum is sufficiently steep, $n\ga 3$, and (b) the inertial range of the initial turbulent velocity field, $\lambda_{_{\rm MAX}}/\lambda_{_{\rm MIN}}=k_{_{\rm MAX}}/k_{_{\rm MIN}}$, is sufficiently large, then $\lambda_{_{\rm MIN}}$ is unimportant, since very little turbulent power is invested in the shortest wavelengths and it is divided between many modes. However, $\lambda_{_{\rm MAX}}$ has a major impact on core collapse, fragmentation, and disk formation. For the low levels of turbulence typical of low-mass cores, dynamical filament fragmentation requires $\lambda_{_{\rm MAX}}\ga R_{_{\rm CORE}}$; fragmentation is very rare when $\lambda_{_{\rm MAX}}\la R_{_{\rm CORE}}/2$. In addition, core angular momenta and the radii of protostellar disks both increase with increasing $\lambda_{_{\rm MAX}}$. These differences arise because large-scale turbulence promotes the formation of large, coherent filaments. Such filaments not only fragment, but also deliver streams of material with disparate specific angular momenta into the center of the core, where this material then forms large disks.

The structure of the paper is as follows. In section 2 we describe the numerical method. We present the results of our simulations in Section 3, discuss the results in section 4, compare with previous work in section 5, and summarize the main conclusions in section 6.

\section{Initial Conditions \& Numerical Method} \label{ICs}

\subsection{Turbulence}

Turbulent cores have irregular internal velocity fields \citep{Belloche2001, Andre2007, Maruta2010}, and these velocities result in a net angular momentum \citep{Goldsmith1985, Dubinski1995, Goodman1993, Barranco1998, Caselli2002}. \citet{Jijina1999} estimate that a low-mass core typically has a ratio of turbulent to gravitational energy in the range $0<\gamma_{_{\rm TURB}}\la 0.5$, and a mean specific angular momentum $j_{_{\rm CORE}}\sim 10^{21}\,{\rm cm}^2\,{\rm s}^{-1}$. \citet{BurkertBodenheimer00} have shown that these features can be reproduced if the turbulence has a power spectrum of the form $P_k \propto k^{-n}$ with $n=3$ or $n=4$, and this finding has been employed by several authors \citep[e.g.][]{Fisher2004, Matzner2005, Kratter2006, Krumholz2007, Kratter2008, Walch2010}. 

Here, we create random, turbulent velocity fields with the same ansatz. First, we generate random Gaussian velocity fields with $P_k\propto k^{-4}$ (Burgers turbulence), populating the wavenumbers $k_{_{\rm MIN}}\leq k\leq k_{_{\rm MAX}}$ in Fourier space. Next, we map the velocity fields onto a uniform $128^3$ grid, and scale them to the required root-mean-square velocity (or equivalently, to the required Mach number, assuming a sound speed of $0.2\,{\rm km}\,{\rm s}^{-1}$). Finally we compute the initial velocities of individual SPH particles by linear interpolation on this grid. This setup results in a ratio of solenoidal to compressive modes close to 2:1.

Previous simulations of turbulent cores have not reported the range of wave-numbers populated in the initial turbulent velocity field. We demonstrate here that this is a fundamental issue. Specifically, the choice of $k_{_{\rm MIN}}$ is critical, but the choice of $k_{_{\rm MAX}}$ is unimportant, provided $k_{_{\rm MAX}}\gg k_{_{\rm MIN}}$. To show this, we consider $k_{_{\rm MIN}}=1/2,\;1,\;2,\;4$, corresponding to $\lambda_{_{\rm MAX}}=2R_{_{\rm CORE}}/k_{_{\rm MIN}}=4R_{_{\rm CORE}},\;2R_{_{\rm CORE}},\;R_{_{\rm CORE}},\;R_{_{\rm CORE}}/2$. For each value of $k_{_{\rm MIN}}$ we simulate five different realizations (by using five different random number seeds 100, 200, 300, 400 and 500), in order to obtain better statistics. In each case we adopt $k_{_{\rm MAX}}=k_{_{\rm MIN}}+8$. Since $n=4$, there is little power on the shorter wavelengths, and so the choice of $k_{_{\rm MAX}}$ is unimportant. 

\subsection{Initial core properties}

We base the initial structure of the simulated core on the observed properties of the A-MM8 core in Ophiuchus.

Using $850\,\mu{\rm m}$ observations, \citet{Simpson2010} estimates the mass to be $M_\mathrm{core}=1.28\,{\rm M}_{_\odot}$, and the azimuthally averaged FWHM diameter to be $D_{_{\rm FWHM}}=2000\,{\rm AU}\;$ (A-MM8 is slightly elongated, with aspect ratio $<1.2$, but we neglect this). The core is modeled with the density profile of a critical Bonnor-Ebert sphere (i.e. the dimensionless boundary radius is $\xi_{_{\rm B}}=6.451$), but it is too cold to be in hydrostatic equilibrium. The outer radius of the core is given by $R_\mathrm{core}=1.25\;{\rm D}_\mathrm{FWHM}=2500\,{\rm AU}$ (on the assumption that the dust temperature is uniform and the $850\,\mu{\rm m}$ emission is optically thin). The central density is $\rho_{_{\rm C}}=6.5\times 10^{-17}\,{\rm g}\,{\rm cm}^{-3}$; and the central free-fall time is $t_{_{\rm FF}}=8.3\,{\rm kyr}$.

Detailed 3D radiative transfer modeling of A-MM8 by Stamatellos et al. (2007) indicates that the mean temperature is $\bar{T}=11\,{\rm K}$, and therefore the ratio of the thermal to gravitational energy is $\alpha_{_{\rm THERM}}=0.017$.

\citet{Andre2007} report a FWHM velocity width for the N$_{_2}$H$^{^+}\;(1-0)$ line of $\Delta v_{_{\rm FWHM}}=0.384\,{\rm km}\,{\rm s}^{-1}$. If the non-thermal velocity dispersion is attributed to turbulence, the turbulent energy is 
\begin{eqnarray}
U_{_{\rm TURB}}&=&\frac{3\,M_\mathrm{core}}{2}\,\left\{\frac{\Delta v_{_{\rm FWHM}}^2}{8\ln(2)}\,-\,\frac{k_{_{\rm B}}\,T_{_{\rm CORE}}}{m_{_{\rm N_2H^+}}}\right\}\,,
\end{eqnarray}
the mean turbulent Mach number is $\bar{\cal M}=0.79$, and the ratio of turbulent to gravitational energy is $\gamma_{_{\rm TURB}}=0.010$.\\

We note that the core mass we are using is $\sim\! 10$ times higher than the value derived by \citet{Andre2007}, who find ${\rm M}_{_{\rm CORE}}=0.13 {\rm M}_\odot$ using 1.2 mm observations and assuming a constant dust temperature of 20 K. On the other hand, it is $\sim 3$ times lower than the mass derived by \citet{Simpson2008}, who find ${\rm M}_{_{\rm CORE}}=3.18 {\rm M}_\odot$. Finally, the virial mass for the given velocity dispersion as derived by \citet{Andre2007} is ${\rm M}_{_{\rm CORE}}=0.71 {\rm M}_\odot$. Given the signifiant spread in observational mass estimates for A-MM8 we have adopted the revised and intermediate value of \citet{Simpson2010}. Our particular choice of parameters results in the core being highly supercritical, with $\alpha_{_{\rm THERM}}+\gamma_{_{\rm TURB}}=0.027$. Hence the results may only pertain to such highly supercritical cores. We will explore the consequences of adopting larger values of $\alpha_{_{\rm THERM}}$ and/or $\gamma_{_{\rm TURB}}$, in a subsequent paper.

\subsection{Gravity and hydrodynamics}

We use the SEREN SPH code \citep{Hubber2011}, which is parallelized using OpenMP and designed for star formation simulations. It has been extensively tested, and applied to a wide range of problems \citep[e.g.][]{Stamatellos2010b, Stamatellos2010, Bisbas2009, Bisbas2010}. It includes both the traditional SPH formulation \citep{Monaghan1992} and the more recent \textit{grad-h} SPH formulation \citep{PriceMonaghan2004}, which we use in this paper. To solve the SPH equations, we employ the symplectic 2nd-order Leapfrog-KDK integrator, in conjunction with a block time-stepping scheme. We invoke additional features within the basic SPH algorithm, such as the Balsara viscosity switch \citep{Balsara1995} to reduce artificial shear viscosity. We use a Barnes-Hut octal-spatial tree \citep{BarnesHut1986} with the GADGET-style multipole acceptance criterion \citep{Springel2001}. Each SPH particle has a mass of $m_{_{\rm SPH}}=10^{-5}\,{\rm M}_{_\odot}$, resulting in a mass resolution of $10^{-3}\,{\rm M}_{_\odot}$ (a Jupiter mass). For random seed 500, we demonstrate convergence by re-simulating the fiducial setup using 10 times as many SPH particles, i.e. with $m_{_{\rm SPH}}=10^{-6}\,{\rm M}_{_\odot}$. These runs are referred to as the '\_hr' simulations. Gravitationally bound condensations that have a central density $\rho_{_{\rm C}}>\rho_{_{\rm SINK}}=10^{-9}\,{\rm g}\,{\rm cm}^{-3}$ are replaced with sinks, and subsequently grow by accretion, using a new algorithm (Hubber, Walch \& Whitworth, 2011) that (a) ensures excellent numerical convergence, and (b) broadcasts the angular momentum of the accreted material to the surrounding gas (rather than assimilating it, which would be non-physical). 

\subsection{Energy equation and radiative transfer}

We use the radiative diffusion approximation of \citet{Stamatellos2007} (RAD-WS method) to solve the energy equation and evaluate radiative transfer effects. The RAD-WS method uses the density, $\rho_{_i}$, temperature, $T_{_i}$, and gravitational potential, $\psi_{_i}$, of an SPH particle $i$ to estimate a mean column-density, $\bar{\Sigma}_i=\Sigma(\rho_{_i},\psi_{_i})$, and a mean optical depth, $\bar{\tau}_{_i}=\tau(\rho_{_i},T_{_i},\psi_{_i})$, through which the particle cools and heats. This optical depth includes contributions from dust, lines and free-free processes, and accounts for the variation in the opacity in the cooler, less dense material that is presumed to surround particle $i$. The net radiative heating rate for the particle $i$ is then
\begin{equation}
\left.\frac{du_i}{dt}\right|_{_{\rm RAD}}=\frac{4\,\sigma_{_{\rm SB}}\,(T_{_{\rm O}}^4 - T_i^4)}
{\bar{\Sigma}_i\,\left\{\bar{\tau}_{_i}+\bar{\tau}_{_i}^{-1}\right\}}\,.
\end{equation}                                                                         
The positive term on the right-hand side represents heating by the background radiation field, and ensures that the gas and dust cannot cool radiatively below the background radiation temperature $T_{_{\rm O}}$, which we set to $T_{_{\rm O}}=7K$. The energy equation then takes account of compressional heating, viscous heating, radiative heating by the background and radiative cooling. It has been extensively tested against detailed numerical \citep{MasunagaInutsuka2000, BossBodenheimer1979, BossMyhill1992, WhitehouseBate2006} and analytical results \citep{Spiegel1957, Hubeny1990}, and performs well in both the optically thin and optically thick regimes.

\section{Results}

\subsection{Core angular momenta}

\begin{figure}
\vspace*{1pt}
\psfig{figure=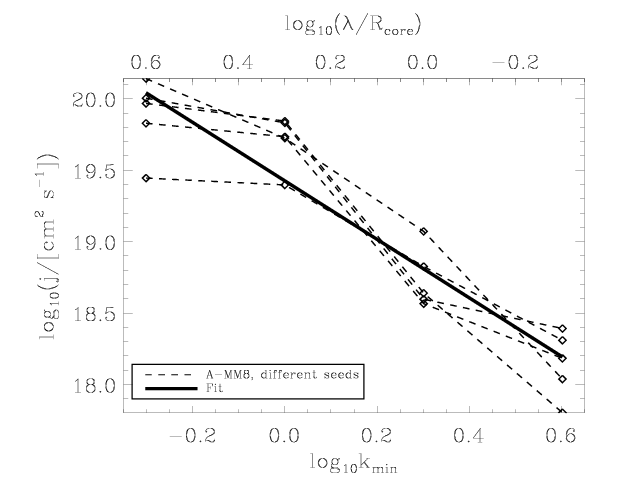 , width=85mm}
\caption{Magnitude of the specific core angular momentum $j_{_{\rm CORE}}$ as a function of $k_{_{\rm MIN}}$ (lower abscissa), or equivalently $\lambda_{_{\rm MAX}}/R_{_{\rm CORE}}$ (upper abscissa). Results obtained with the same seed but different $k_{_{\rm MIN}}$ are connected with dashed lines. The thick black line is the best fit to the data in the interval $1\la k_{_{\rm MIN}}\la 4$ (see Eqn. \ref{EQN:jFIT}).}
\label{FIG1}
\end{figure}

As $k_{_{\rm MIN}}$ increases, the coherence lengths of the most energetic turbulent modes ($\la\!\lambda_{_{\rm MAX}}/2\!\sim\!R_{_{\rm CORE}}/k_{_{\rm MIN}}$) decrease. Since these modes are uncorrelated, the specific angular momentum, 
\begin{eqnarray}
{\bf j}_{_{\rm CORE}}&=&\frac{1}{{\cal N}_{_{\rm SPH}}}\,\sum\limits_{i=1}^{{\cal N}_{_{\rm SPH}}}\,\left\{{\bf r}_{_i}\times{\bf v}_{_i}\right\}\,,
\end{eqnarray}
which is compounded by contributions from all the different modes, also decreases in magnitude. Fig. \ref{FIG1} shows the variation of $j_{_{\rm CORE}}$ with $k_{_{\rm MIN}}$ at $t=0$. For each value of $k_{_{\rm MIN}}$ we simulate five different realizations by invoking five different seeds, and this produces an $\sim\! 0.5\,{\rm dex}$ spread in $j_{_{\rm CORE}}$. However, there is an underlying systematic variation that, in the interval $1\la k_{_{\rm MIN}}\la 4$, can be approximated by
\begin{eqnarray}\label{EQN:jFIT}
j_{_{\rm CORE}}&\simeq&4\times 10^{19}\,{\rm cm}^2\,{\rm s}^{-1}\,k_{_{\rm MIN}}^{-2}\,.
\end{eqnarray}
At smaller $k_{_{\rm MIN}}$, $\;j_{_{\rm CORE}}$ approaches the maximum value consistent with the amount of turbulent energy (see below). 
In addition, much of the turbulent energy is invested in modes of such long wavelength that it results in bulk motion of the core, rather than intrinsic spin. The overall range of values is $10^{17.8}\,{\rm cm}^2\,{\rm s}^{-1}\la j_{_{\rm CORE}}\la 10^{20.2}\,{\rm cm}^2\,{\rm s}^{-1}$.

We note that for a critical Bonnor-Ebert sphere, the ratio of rotational to gravitational energy is given by $\beta_{_{\rm ROT}}=(j_{_{\rm CORE}}/j_{_\beta})^2$, where $j_{_\beta}=0.644(GM_{_{\rm CORE}}R_{_{\rm CORE}})^{1/2}$. For A-MM8, $j_{_\beta}=1.6\times 10^{21}\,{\rm cm}^2\,{\rm s}^{-1}$ and hence
\begin{eqnarray}
\beta_{_{\rm ROT}}&=&\left(\frac{j_{_{\rm CORE}}}{1.6\times 10^{21}\,{\rm cm}^2\,{\rm s}^{-1}}\right)^2\,.
\end{eqnarray}
Since $\beta_{_{\rm ROT}}\leq\gamma_{_{\rm TURB}}=0.01$ (the rotational energy cannot exceed the turbulent energy), there is a maximum specific angular momentum $j_{_{\rm MAX}}=\gamma_{_{\rm TURB}}^{1/2}j_{_\beta}=1.6\times 10^{20}\,{\rm cm}^2\,{\rm s}^{-1}$. Evidently this rather small $j_{_{\rm MAX}}$ is a consequence of small $R_{_{\rm CORE}}$ (observational estimates of $j$ tend to derive from more extended cores) and small $\gamma_{_{\rm TURB}}$.

\subsection{Disk density distributions}

At $t_{_{50}}$, the time at which $50\%$ of the initial core mass has been converted into protostars, we define a Cartesian frame of reference, $(x_{_{\rm IF}},y_{_{\rm IF}},z_{_{\rm IF}})$, in which the $z_{_{\rm IF}}$-axis is aligned with the largest principal moment of inertia of the remaining dense $(\rho>10^{-12}\,{\rm g}\,{\rm cm}^{-3})$ material. The algorithm for doing this is described in Appendix A. Provided there is a single dominant primary disk, $z_{_{\rm IF}}$ is then aligned with its rotation axis. Fig. 2 displays a montage of false-color images of the density on the $z_{_{\rm IF}}=0$ plane for the entire ensemble of simulations of A-MM8. Each column of images corresponds to a different value of $k_{_{\rm MIN}}$, and each row to a different seed. From these plots we see that (a) the sizes of disks tend to decrease with increasing $k_{_{\rm MIN}}$; and (b) that for $k_{_{\rm MIN}}=1\;{\rm and}\;2$, multiple protostars usually form, whereas for $k_{_{\rm MIN}}=1/2\;{\rm and}\;4$, only single stars are formed. 

\begin{figure*}
\begin{centering}
\begin{tabular}[t]{c c}
 &   $k_\mathrm{MIN}=1/2$ \hspace{1.8cm} $k_\mathrm{MIN}=1$ \hspace{1.8cm} $k_\mathrm{MIN}=2$ \hspace{1.8cm} $k_\mathrm{MIN}=4$ \\
 & $\lambda_\mathrm{MAX}/R_{_{\rm CORE}}=4$ \hspace{0.8cm} $\lambda_\mathrm{MAX}/R_{_{\rm CORE}}=2$ \hspace{0.8cm} $\lambda_\mathrm{MAX}/R_{_{\rm CORE}}=1$ \hspace{0.8cm} $\lambda_\mathrm{MAX}/R_{_{\rm CORE}}=1/2$\\
    seed 200 & 
    \psfig{figure=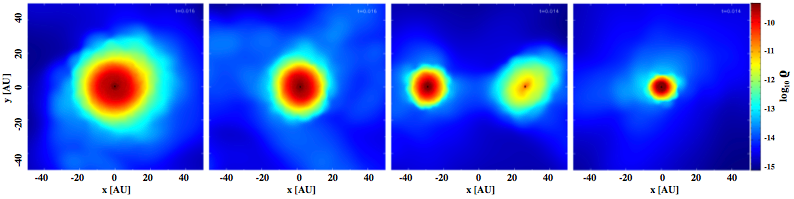, width=158mm}  \\
    seed 300 &
    \psfig{figure=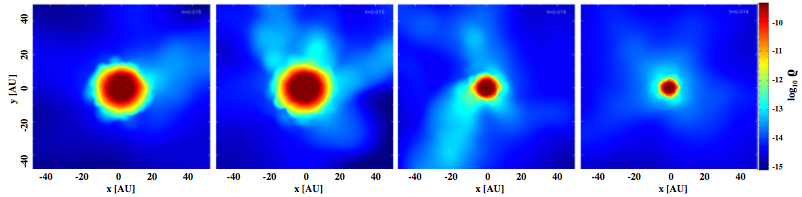, width=158mm}  \\
    seed 400 &
    \psfig{figure=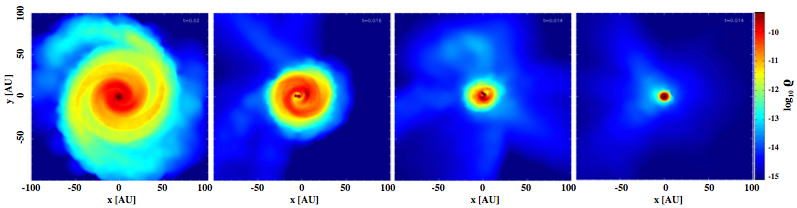, width=158mm}  \\
    seed 500 &
    \psfig{figure=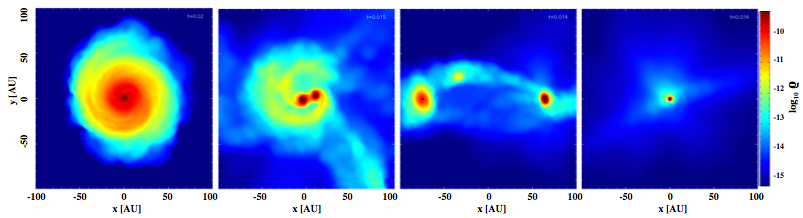, width=158mm}  \\
    seed 600 &
    \psfig{figure=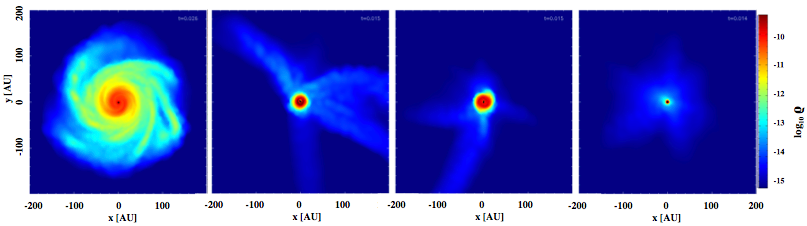, width=158mm}  \\
\end{tabular}
\caption{Montage of false-color images of the density on the midplanes of the protostellar disks formed in the entire ensemble of simulations of A-MM8. Each column corresponds to a different value of $k_{_{\rm MIN}}$ (from left to right: $1/2$, $1$, $2$, and $4$), and hence to a different $\lambda_{_{\rm MAX}}=2R_{_{\rm CORE}}/k_{_{\rm MIN}}$ (from left to right: $4R_{_{\rm CORE}}$, $2R_{_{\rm CORE}}$, $R_{_{\rm CORE}}$, and $R_{_{\rm CORE}}/2$). Each row corresponds to a different seed for generating the initial turbulent velocity field. The false-color encodes the same range of density on all plots: $10^{-15}\,{\rm g}\,{\rm cm}^{-3}$ to $10^{-11}\,{\rm g}\,{\rm cm}^{-3}$. However, the linear sizes of the frames are different for different seeds, varying from $100\,{\rm AU}$ to $400\,{\rm AU}$. Black dots mark the positions of sink particles.\label{FIG2A}}
\end{centering}
\end{figure*}

\begin{table}
\begin{center}
\begin{tabular}{cccccl}
$k_{_{\rm MIN}}$ & {\sc seed} & $t_{_{\rm 50}}$      & $M_{_{\rm DISK}}$             & $R_{_{\rm DISK}}$   & $M_{_\star}$                  \\
                 &            & $\overline{\rm kyr}$ & $\overline{{\rm M}_{_\odot}}$ & $\overline{\rm AU}$ & $\overline{{\rm M}_{_\odot}}$ \\\hline
1/2              & 200        & 16.2                 & 0.39                          & 30                  & .640                          \\
                 & 300        & 14.7                 & 0.26                          & 17                  & .640                          \\
                 & 400        & 20.2                 & 0.57                          & 94                  & .640                          \\
                & 500        & 20.3                 & 0.53                          & 47                  & .640                          \\
                & 500\_hr   & 16.0              &  0.59                        &  45                   & .41     \\
                 & 600        & 26.0                 & 0.59                          & 100                 & .640                          \\\hline
1                & 200        & 16.2                 & 0.27                          & 16                  & .628/.012                     \\
                 & 300        & 15.3                 & 0.30                          & 16                  & .640                          \\
                 & 400        & 15.9                 & 0.37                          & 40                  & .430/.210                     \\
                & 500        & 15.2                 & 0.31                          & 30                  & .360/.270/.010                \\
               & 500\_hr    &  14.2          &  0.14            &  30           & .323/.318                         \\
               & 600        & 15.3                 & 0.26                          & 25                  & .310/.190/.140                \\\hline
2                & 200        & 14.3                 & 0.10                          & 15                  & .588/.052                     \\
                 & 300        & 14.6                 & 0.15                          & $\;\,8$             & .320/.320                     \\
                 & 400        & 14.1                 & 0.14                          & 17                  & .330/.163/.147                \\
                 & 500        & 14.2                 & 0.19                          & 15                  & .616/.024    \\
                 & 600        & 14.6                 & 0.25                          & 20                  & .250/.207/.183                \\\hline
4                & 200        & 13.9                 & 0.05                          & $\;\,8$             & .640                          \\
                 & 300        & 13.8                 & 0.10                          & $\;\,6$             & .640                          \\
                 & 400        & 13.9                 & 0.12                          & $\;\,9$             & .640                          \\
                 & 500        & 13.8                 & 0.10                          & $\;\,9$             & .640                          \\
                & 500\_hr   &  13.7              & 0.16                           & $\;\;9$             & .640                          \\
                 & 600        & 13.8                 & 0.11                          & $\;\,6$             & .640                          \\\hline
\end{tabular}
\caption{Properties of protostellar disks and protostars. Column 1 gives $k_{_{\rm MIN}}$. Column 2 gives the seed used to generate the initial turbulent velocity field. The extension '\_hr' denotes the high-resolution runs. Column 3 gives the time at which half the initial core mass has been converted into protostars, $t_{_{50}}$, in kyr. Column 4 gives the disk mass, $M_{_{\rm DISK}}$, in ${\rm M}_{_\odot}$. Column 5 gives the disk radius, $R_{_{\rm DISK}}$, in AU. Column 6 gives the masses of the protostars formed, $M_{_\star}$, in ${\rm M}_{_\odot}$.}
\end{center}
\label{TAB:PROPS}
\end{table}

\subsection{Resolution study}
\begin{figure*}
\begin{centering}
\begin{tabular}[t]{c c}
 &   $k_\mathrm{MIN}=1/2$ \hspace{1.8cm} $k_\mathrm{MIN}=1$ \hspace{1.8cm} $k_\mathrm{MIN}=2$ \hspace{1.8cm} $k_\mathrm{MIN}=4$ \\
 & $\lambda_\mathrm{MAX}/R_{_{\rm CORE}}=4$ \hspace{0.8cm} $\lambda_\mathrm{MAX}/R_{_{\rm CORE}}=2$ \hspace{0.8cm} $\lambda_\mathrm{MAX}/R_{_{\rm CORE}}=1$ \hspace{0.8cm} $\lambda_\mathrm{MAX}/R_{_{\rm CORE}}=1/2$\\
     & 
    \psfig{figure=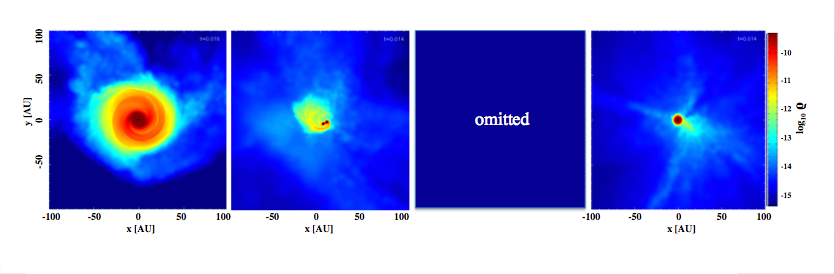, width=158mm}  \\
\end{tabular}
\caption{Same as Fig. 2 but for the high resolution runs with seed 500 ({\it 500\_hr}). \label{FIG2HR}}
\end{centering}
\end{figure*}

We re-simulate three setups for seed 500 ($k_{_{\rm MIN}}=1/2,\; 1, \;{\rm and}\; 4$) with ten times higher resolution (run {\it 500\_hr}), i.e. a total of 1,280,000 particles and $m_{_{\rm SPH}}=10^{-6}$.
The results of these simulations are shown in Fig. \ref{FIG2HR}. With regard to the density profiles and the disk radii, we find remarkably good convergence between low and high resolution runs (see also Table \ref{TAB:PROPS}). The disk masses are also in reasonable agreement, considering the fact that the end times are slightly different. For $k_{_{\rm MIN}}=1/2$ we could not follow the simulation until 50\% of the core mass collapsed into the sink because of CPU time limitations. Therefore, the runs are mistimed and the disk masses cannot be strictly compared in this case.  For $k_{_{\rm MIN}}=1$, the sink masses grow a bit quicker in the high resolution run ($t_{_{\rm 50}}=14.2$ kyr instead of $t_{_{\rm 50}}=15.2$ kyr) leading to a smaller disk mass of $0.14 M_\odot$ rather than $0.31 M_\odot$. This difference may be caused by the different fragmentation properties as in run  {\it 500\_hr}  only two instead of three sink particles form. Despite the different number of sink particles, the disk density distributions and the 'system masses' within the two identifiable, individual condensations (see Fig. 2 and Fig. \ref{FIG2HR}) are very similar, i.e. $M_{_\star}=0.323\; {\rm and}\; 0.318$ for {\it 500\_hr} and $M_{_\star}=0.36\; {\rm and}\; (0.27 + 0.01)=0.28$ for {\it 500}. For $k_{_{\rm MIN}}=4$, we find good agreement of all interesting quantities. Overall, the results of our study are only weakly dependent on resolution.

\subsection{Global disk properties and scaling relations}

\begin{figure*} \label{FIG3}
\begin{centering}
\vspace*{1pt}
\begin{tabular}[t]{c c}
    \psfig{figure=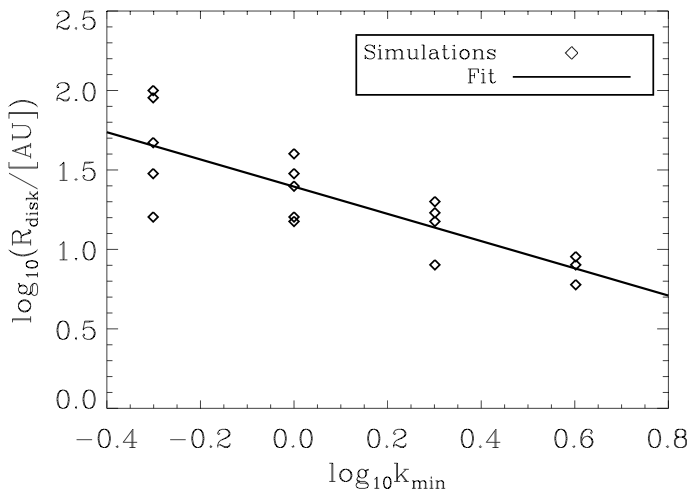, width=85mm} & \psfig{figure=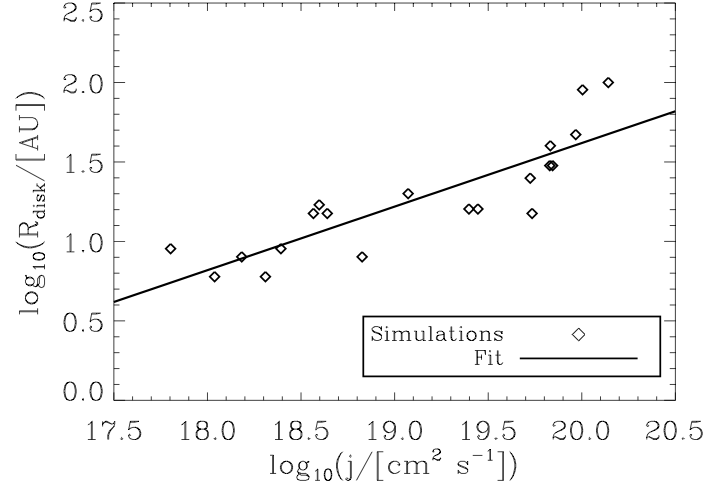, width=85mm} \\
    \psfig{figure=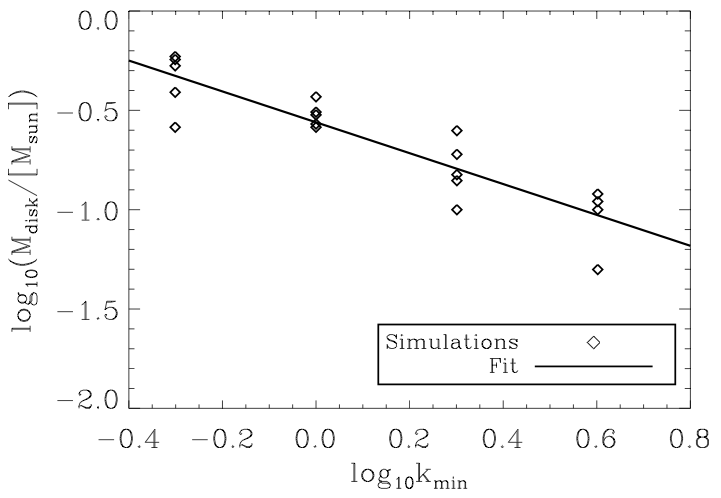, width=85mm} & \psfig{figure=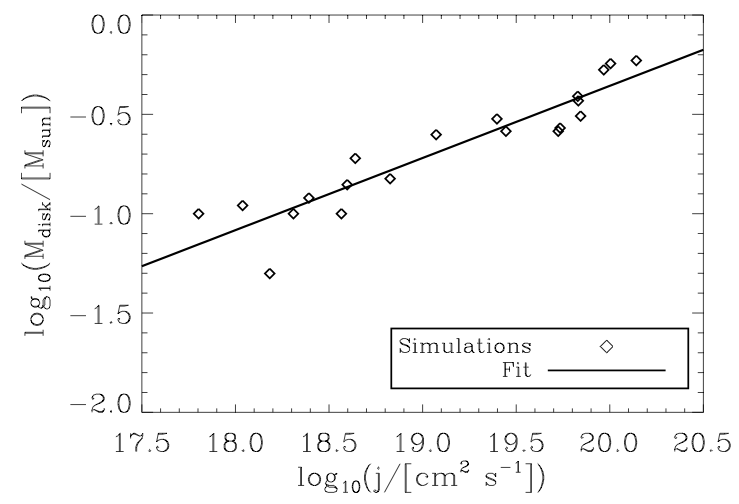, width=85mm} \\
\end{tabular}
\caption{Disk radius, $R_{_{\rm DISK}}$, and disk mass, $M_{_{\rm DISK}}$, as a function of minimum turbulent wavenumber, $k_{_{\rm MIN}}$, and specific angular momentum $j_{_{\rm CORE}}$.}
\end{centering}
\end{figure*}

\begin{figure}\label{FIG4}
\begin{centering}
\vspace*{1pt}
  \psfig{figure=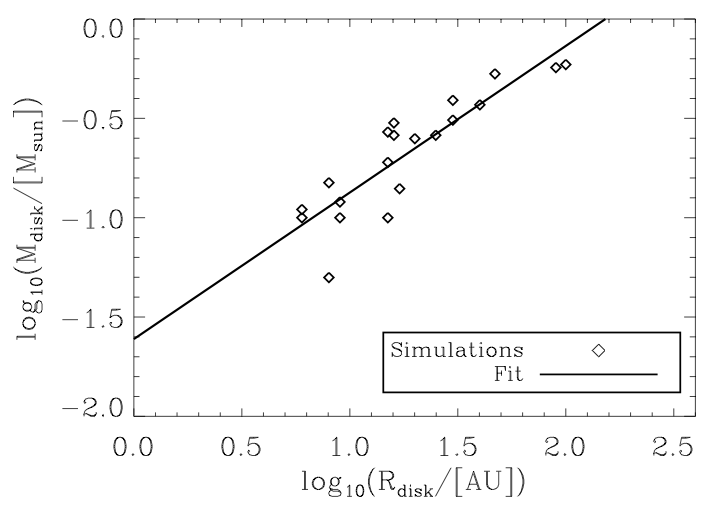, width=80mm}
\caption{The correlation of disk mass, $M_{_{\rm DISK}}$, with disk radius, $R_{_{\rm DISK}}$.} 
\end{centering}
\end{figure}

There are many types of disk: circumstellar, circum-binary and circum-system (i.e. enclosing higher order systems). Here we focus our discussion on the primary accretion disk, which is the most massive and extended disk in the simulation (at $t_{_{50}}$), and always surrounds the protostar with the highest final mass -- although this is not always the first protostar to form. The typical primary disk is rather small, $R_{_{\rm DISK}}\sim 10\;{\rm to}\;30\,{\rm AU}$, and has mass $M_{_{\rm DISK}}\sim 0.1\;{\rm to}\;0.6\,{\rm M}_{_\odot}$. In general more extended disks are more massive, and the most extended disks display signs of  gravitational instability. However, this only results in the formation of spiral arms and the transport of angular momentum; there are no protostars formed by disk fragmentation. Table 1 lists for each simulation, the value of $k_{_{\rm MIN}}$, the seed used to generate the turbulent velocity field, the time to convert half the core mass into stars, $t_{_{50}}$, the mass and radius of the primary disk at this time, $M_{_{\rm DISK}}$ and $R_{_{\rm DISK}}$, and the masses of the protostars formed. Fig. 3 shows the dependencies of $R_{_{\rm DISK}}$ and $M_{_{\rm DISK}}$ on $k_{_{\rm MIN}}$ and $j_{_{\rm CORE}}$; and Fig. 4 shows the correlation between $M_{_{\rm DISK}}$ and $R_{_{\rm DISK}}$. The linear fits on these figures, and their uncertainties, obtained by $\chi^2$ minimization, are
\newpage
\begin{eqnarray}
R_{_{\rm DISK}}\!&\!\simeq\!&\!26(\pm2)\,{\rm AU}\;\,k_{_{\rm MIN}}^{-0.86(\pm0.13)}\,,\\
R_{_{\rm DISK}}\!&\!\simeq\!&\!16(\pm2)\,{\rm AU}\;\left(\frac{j_{_{\rm CORE}}}{10^{19}\,{\rm cm}^2\,{\rm s}^{-1}}\right)^{0.40(\pm0.06)}\,,\\\nonumber&&\\
M_{_{\rm DISK}}\!&\!\simeq\!&\!0.28(\pm0.02)\,{\rm M}_{_\odot}\;\,k_{_{\rm MIN}}^{-0.73(\pm0.09)}\,,\\
M_{_{\rm DISK}}\!&\!\simeq\!&\!0.19(\pm0.01)\,{\rm M}_{_\odot}\;\left(\frac{j_{_{\rm CORE}}}{10^{19}\,{\rm cm}^2\,{\rm s}^{-1}}\right)^{0.36(\pm0.03)}\,,\\\nonumber&&\\
M_{_{\rm DISK}}\!&\!\simeq\!&\!0.30(\pm0.03)\,{\rm M}_{_\odot}\;\left(\frac{R_{_{\rm DISK}}}{30\,{\rm AU}}\right)^{0.74(\pm0.09)}\,.
\end{eqnarray}

\section{Discussion}

\subsection{Filament formation}

\begin{figure*}
\begin{centering}
\vspace*{1pt}
\psfig{figure=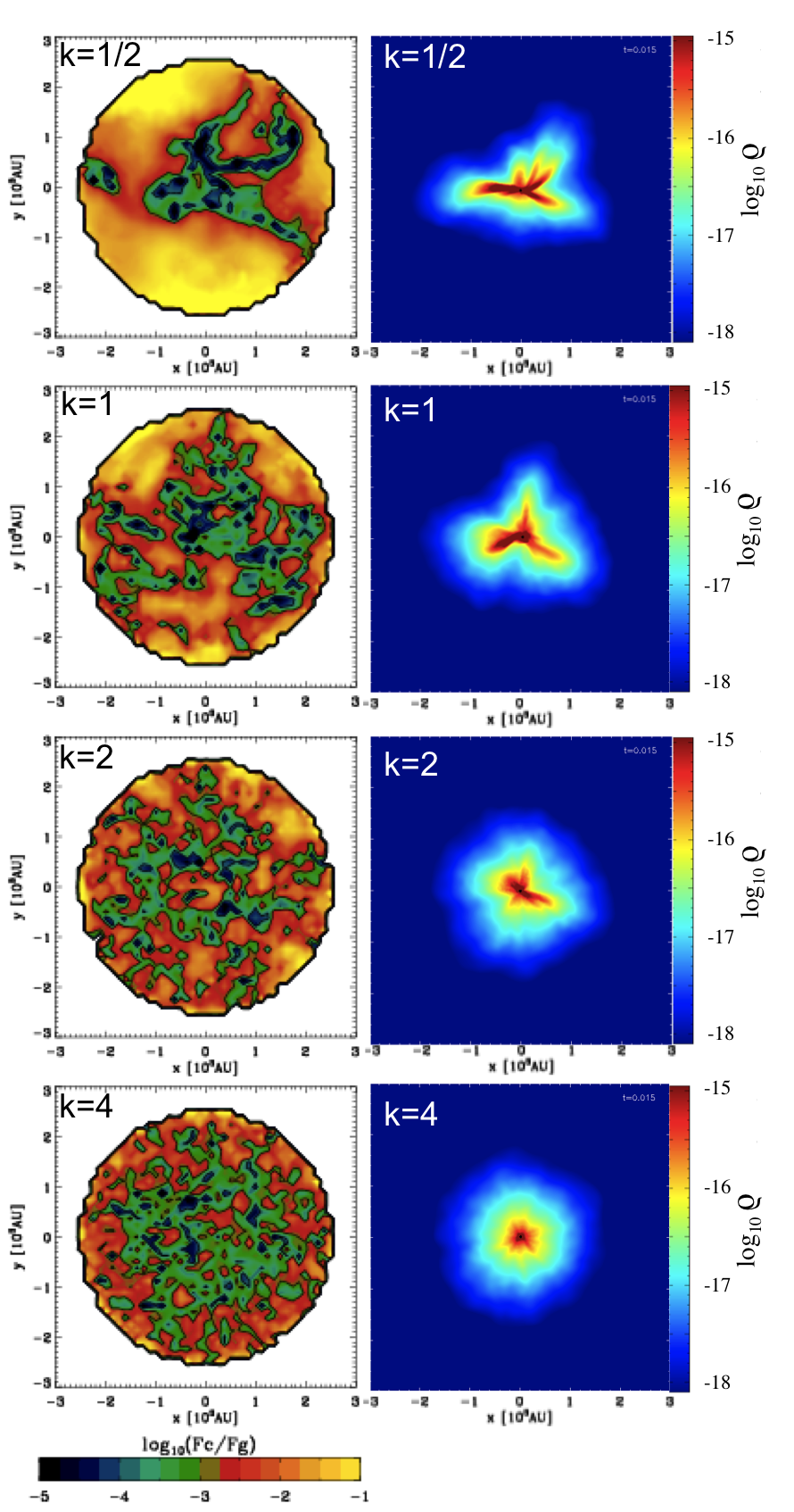, width=115mm} 
\caption{{\sc Left column.} The minimum value of the ratio of centrifugal to gravitational potential force, $F_{\rm C}^i/F_{\rm G}^i$, found along the line of sight, for the initial conditions of the simulations with seed 200 and different $k_{_{\rm MIN}}$. For $k_{_{\rm MIN}}=1/2$ we find coherent features, whereas the distribution of small $F_{\rm C}^i/F_{\rm G}^i$ becomes increasingly random with increasing $k_{_{\rm MIN}}$. {\sc Right column.} The maximum density along the line of sight after $15\,{\rm kyr}$ of evolution for the simulations with seed 200 and different $k_{_{\rm MIN}}$. Filaments have formed where $F_{\rm C}^i/F_{\rm G}^i$ was initially small.}
\label{FIG_FCSD}
\end{centering}
\end{figure*}

Large-scale filaments play a critical role in core fragmentation, and in the formation of the primary disk. First, large-scale filaments provide alternative sites (alternative to the center of mass of the core) where material can converge and form secondary protostars. Second, large-scale filaments deliver large parcels of material with disparate angular momenta into the center, where they accumulate in a large disk around the primary protostar. Figs. \ref{FIG_FCSD} demonstrates that large-scale filaments are generated by turbulence with small $k_{_{\rm MIN}}$.

The right-hand column of Fig. \ref{FIG_FCSD} shows false-color images of the maximum density on lines of sight parallel to the $z_{_{\rm IF}}$ axis, at time $t=15\,{\rm kyr}$, for different values of $k_{_{\rm MIN}}$. All the simulations presented in Fig. \ref{FIG_FCSD} derive from the same seed, but the results obtained with other seeds are statistically similar. We see that for $k_{_{\rm MIN}}=1/2$ the high-density gas ($n\ga 10^6\,{\rm H}_{_2}\,{\rm cm}^{-3}$) is concentrated in large-scale filaments. However, as $k_{_{\rm MIN}}$ is increased, the strength and coherence of the filaments declines, and by $k_{_{\rm MIN}}=4$ there are no noticeable filaments. 
\label{FIG2A}
Filament formation can be understood in terms of the forces shaping the core. We neglect the pressure force, $F_{\rm P}$, since $F_{\rm P}$ is initially smooth, and focus on the ratio of centrifugal to gravitational force $F_{\rm C}/F_{\rm G}$. For each SPH particle, $i$, we compute $F_{\rm C}^i=|{\bf r}_i\!\stackrel{\,}{\wedge}\!{\bf v}_i|^2/|{\bf r}_i|^3$ and $F_{\rm G}^i=GM(|{\bf r}_i|)/|{\bf r}_i|^2$, where ${\bf r}_i$ and ${\bf v}_i$ are the position and velocity of particle $i$ relative to the centre of mass, and $M(|{\bf r}_i|)$ is the mass interior to radius $|{\bf r}_i|$. Parcels of gas with low centrifugal support, i.e. small $F_{\rm C}/F_{\rm G}$, collapse first, and neighboring parcels are then drawn into the regions they vacate, creating preferred accretion streams, i.e. filaments. The left-hand column of \ref{FIG_FCSD} shows false color images of the minimum value of $F_{\rm C}^i/F_{\rm G}^i$ found on each line of sight. For $k_{_{\rm MIN}}=1/2$, there are well defined structures with low $F_{\rm C}/F_{\rm G}$ that can be related to the filaments illustrated in the corresponding right-hand image. However, as $k_{_{\rm MIN}}$ is increased, structures with low $F_{\rm C}/F_{\rm G}$ become increasingly small and incoherent.

A complex of filamentary structures on scales of 1000 AU, very similar to our case with small $k_{_{\rm MIN}}$, has recently been observed in the envelopes of Class 0 cores by \citet{Tobin2010} using {\sc Spitzer}. \citet{Tobin2010} note that this complex envelope structure is spatially distinct from possible outflow cavities, and explicitly suggest that it results from the collapse of prestellar cores with initial non-equilibrium structures.

\subsection{Protostellar multiplicity}

\begin{figure}
\begin{centering}
\psfig{figure=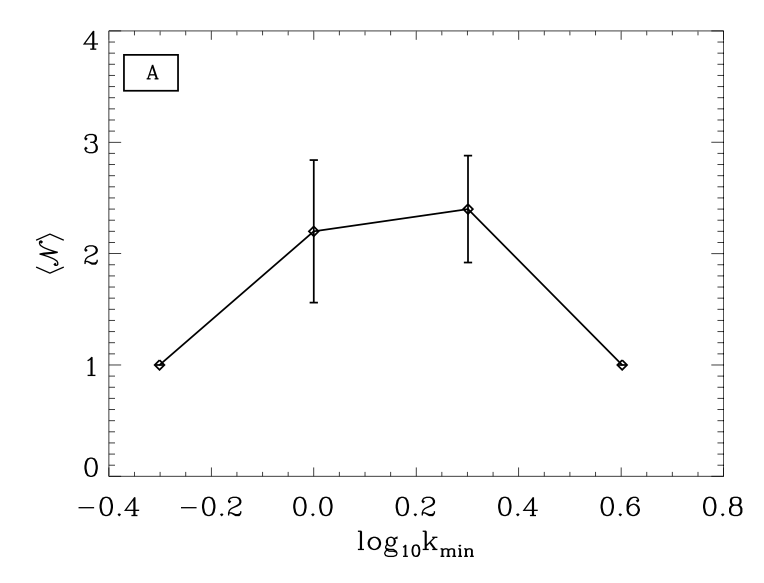, width=80mm} 
\psfig{figure=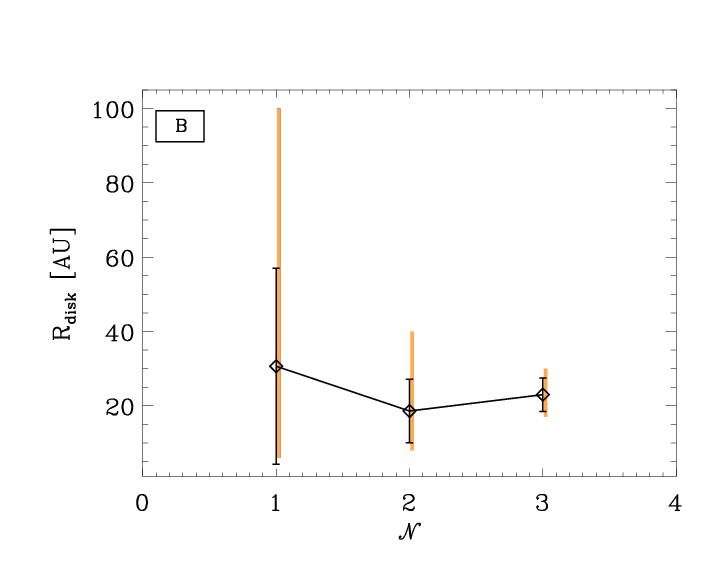, width=80mm}
\caption{{\sc Panel A:} Mean (diamond) and variance (error bar) of the number of protostars formed, as a function of $k_\mathrm{MIN}$. {\sc Panel B:} Mean (diamond), variance (error bar), and range (mandarine line) of the disk radii at $t_{_{50}}$ as a function of the number of protostars formed.}
\label{FIG_jkmulti}
\end{centering}
\end{figure}

From Table 1 it appears that cores with $k_{_{\rm MIN}}=1/2\;{\rm and}\;4$ only spawn single stars, whereas cores with $k_{_{\rm MIN}}=1\;{\rm and}\;2$ tend to spawn 2 or 3 stars. In other words, multiple systems are formed only if $\lambda_{_{\rm MAX}}/R_{_{\rm CORE}}\sim 1\;{\rm to}\;2$. 

This is illustrated in {\sc Panel A} of Fig. \ref{FIG_jkmulti}, where we plot the mean number of stars formed, and its variance, against $k_{_{\rm MIN}}$.
There are no instances of disk fragmentation. In all cases where secondary stars form they form by dynamical filament fragmentation. Even where two (or three) stars end up in a close binary (triple) system with a circum-binary (circum-system) disk, the components have always formed by filament fragmentation, with their own independent circumstellar disks, and then fallen into the center and captured one another.  This result was already suggested by the results of Walch et al. (2010), but the current study places it on a quantitative footing. 

{\sc Panel B} of Fig. 6, shows that by $t_{_{50}}$ there is a wide range of disk sizes around single stars, but the disks around stars in multiple systems have been truncated by mutual tidal interactions.
When $k_{_{\rm MIN}}\!=\!1/2$, the filaments do not fragment because they fall into the center very rapidly and consequently the material in them is stretched. On arrival in the center, much of this material is initially parked in a massive extended disk around the primary protostar. However, despite being massive and extended, this disk does not fragment. Repeated perturbations due to irregular infall from the filaments (a) maintain a relatively high velocity dispersion in the disk (and hence a high Toomre ${\cal Q}$ parameter), (b) excite density waves that transport angular momentum by gravitational torques, thereby facilitating accretion onto the central primary star, and (c) shear proto-condensations apart. In a similar vein, \citet{Hayfield2010} have recently shown that disks in binary systems tend to be more stable towards fragmentation than disks around single stars. 

When $k_{_{\rm MIN}}\!=\!4$, there are no significant filaments, so dynamical fragmentation is suppressed. The disks that form around the primary protostar are too small to fragment.

We stress that these results are for a specific low-mass, low-turbulence core. The critical value of $k_\mathrm{MIN}$, above which fragmentation is inhibited, probably increases with increasing core mass, since a small low-contrast filament is more likely to fragment if it is more massive. Also, we might expect more fragmentation for increased levels of turbulence.

\section{Comparison with previous work}

There have been several other studies of the collapse and fragmentation of low-mass, turbulent cores.

\citeauthor{Goodwin2004a} (\citeyear{Goodwin2004a, Goodwin2004b}) simulate the collapse and fragmentation of cores having mass $M_{_{\rm CORE}}\!=\!5.4\,{\rm M}_{_\odot}$ and radius $R_{_{\rm CORE}}\!=\!50,000\,{\rm AU}$, modeled with a Plummer-like density profile. These are SPH simulations, using a barotropic equation of state and sink particles. Different levels of turbulence are considered, $0.01\leq\gamma_{_{\rm TURB}}\leq 0.25$, with $P_{_k}\!\propto\! k^{-4}$, $k_{_{\rm MIN}}\!=\!1$ and purely solenoidal modes; for each case many realizations are performed. In contrast to the simulations presented here (which result in fragmentation for $\gamma_{_{\rm TURB}}=0.01$, provided $1\!\la\!k_{_{\rm MIN}}\!\la\!2$), Goodwin et al. obtain fragmentation only when $\gamma_{_{\rm TURB}}\!\ga\!0.05$. This is due to the fact that their cores have a much higher level of thermal support than ours, $0.30\!\la\!\alpha_{_{\rm THERM}}\!\la\!0.45$, so the gas is less readily compressed, and to the fact that solenoidal modes produce less compression than the thermal mix of solenoidal and compressive modes that we use.

In a related study \citet{Goodwin2006} extend the study with $\gamma_{_{\rm TURB}}=0.10$ to different turbulent power spectra, $P_{_k}\propto\!k^{-n}$ with $n=3,\;4,\;5$. They find that with higher $n$ (i.e. a higher concentration of power at long wavelengths), there is more fragmentation, and the protostars formed have somewhat lower masses.

These simulations have been repeated by \citet{Attwood2009}, with the same initial conditions, but solving the energy equation and treating the associated transport of cooling radiation, instead of using a barotropic equation of state. The main differences in the results are (i) that fragmentation is more efficient (larger numbers of protostars are formed), and (ii) that the binary systems have shorter periods, higher eccentricities, and smaller mass-ratios.

\citet{Walch2010} use SPH to simulate the collapse and fragmentation of cores having mass $M_{_{\rm CORE}}\!=\!6.1\,{\rm M}_{_\odot}$ and radius $R_{_{\rm CORE}}\!=\!17,000\,{\rm AU}$, modeled as marginally supercritical Bonnor-Ebert spheres ($\xi_{_{\rm B}}=6.9$, density increased by $10\%$). The cores have an initial ratio of thermal to gravitational energy $\alpha_{_{\rm THERM}}\!=\!0.74$, and are contained by an external pressure $P_{_{\rm EXT}}\!=\!9\times 10^{-11}\,{\rm erg}\,{\rm cm}^{-3}$. The turbulent velocity field is characterized by a mean Mach number ${\cal M}=1$ (i.e. transsonic turbulence) a power spectrum $P_{_k}\!\propto\!k^{-4}$, $\;k_{_{\rm MIN}}\!=\!1$, and a thermal mix of solenoidal and compressive modes. A large ensemble of cores is generated, and from these a representative subset, having specific angular momenta spanning the range $0.1\!\la\!(j_{_{\rm CORE}}/10^{21}\,{\rm cm}^2\,{\rm s}^{-1})\!\la\!2.7$, is extracted and evolved. The energy equation is solved using the molecular-line cooling rates of \citet{Neufeld1995}. As in the simulations presented here, Walch et al. find that dynamical filament fragmentation dominates over disk fragmentation. However, the gas in their simulations is much hotter (because dust cooling is not included), so the disks that form are more swollen. In addition, since they do not use sink particles, they are unable to follow the simulations to the point where multiple protostars with circumstellar, circum-binary and circum-system disks are formed.

\citet{Offner2008} simulate the collapse and fragmentation of turbulent cores, using AMR and a barotropic equation of state. Their cores are produced in a large-scale simulation of a collapsing molecular cloud, with either driven or decaying turbulence. Individual cores are then followed at higher resolution. \citet{Offner2008} find that simulations with decaying turbulence form on average more low-mass protostars than simulations with driven turbulence. \citet{Offner2009} simulate fragmentation in a turbulent box, using AMR and solving the energy equation. They show that radiative feedback from the forming protostars inhibits disk fragmentation, thereby reducing the number of low-mass multiple systems formed \citep[see also][]{Bate2009c}. Further analysis of these results \citep{Offner2010} suggests that dynamical filament fragmentation is the dominant mechanism  forming low-mass stars and binary systems, rather than disk fragmentation. The material in filaments is sufficiently far from protostellar radiation sources to keep cool and fragment, whereas the material in disks around newly formed protostars is close and gets heated up so that it does not fragment. 

However, \citet{Stamatellos2011} show that, if accretion onto a protostar is episodic (as is believed to be the case), the luminosity is also episodic, and the duty cycle has sufficiently long low-luminosity periods for the outer parts of a massive accretion disk to cool down and fragment. Disk fragmentation may therefore still be a viable mechanism for forming brown dwarfs.


\section{Conclusions}

We have performed an ensemble of SPH self-gravitating radiation-hydrodynamic simulations to demonstrate that -- in low-mass turbulent cores -- the largest wavelength in the turbulent spectrum has a critical bearing on the outcome of collapse and fragmentation. Specifically, if all other parameters (the initial critical Bonnor-Ebert density profile, $\alpha_{_{\rm THERM}}\!=\!0.017$, $\gamma_{_{\rm TURB}}\!=\!0.010$, $n\equiv -d\ln[P_{_k}]/d\ln[k]\!=\!4$) are held fixed and $\lambda_{_{\rm MAX}}$ is varied,

\begin{itemize}

\item{the mean specific angular momentum of the core increases approximately as the square of the the largest wavelength, $j_{_{\rm CORE}}\sim\lambda_{_{\rm MAX}}^2,\;$ for $1/2\!\la\!\lambda_{_{\rm MAX}}/R_{_{\rm CORE}}\!\la\!2\,$;}

\item{filaments form in the regions where the centrifugal support is weakest, and therefore the material collapses fastest;}

\item{the size and coherence of filaments therefore increases with increasing $\lambda_{_{\rm MAX}}$;}

\item{dynamical filament fragmentation \citep[cf.][]{Offner2010} is the dominant (only) fragmentation mechanism,}

\item{and hence fragmentation and multiple star formation only occur for $\;1/2\!\la\!\lambda_{_{\rm MAX}}/R_{_{\rm CORE}}\!\la\!2\,$;}

\item{the primary (i.e. most massive) protostellar disk has mass and radius which scale approximately as $M_{_{\rm DISK}}\sim\lambda_{_{\rm MAX}}^{3/4}$ and $R_{_{\rm DISK}}\sim\lambda_{_{\rm MAX}}\,$;}

\item{massive extended disks form (for large $\lambda_{_{\rm MAX}}$) where the filamentary inflows deliver material with disparate specific angular momentum;}

\item{but these disks do not fragment, because the inflowing material maintains a large velocity dispersion and therefore the gravitational modes excited in the disk are only strong enough to redistribute angular momentum and facilitate accretion onto the central protostar.}

\end{itemize}

The global parameters of a core do not completely specify the initial conditions for a simulation. In particular, the initial turbulent velocity field is stochastic. Consequently there is considerable variance amongst different realizations of the same parameter set, and the conclusions listed above should be interpreted as statistical.

\section*{Acknowledgments}

We thank the anonymous referee for a thorough and constructive report which helped us to improve the original version of this paper. We acknowledge the support of the Marie Curie RTN {\sc 'CONSTELLATION'} (MRTN-CT-2006-035890). AW further acknowledges the support of Grant ST/HH001530/1 from the UK Science and Technology Facilities Council. The simulations have been carried out on the {\sc ARCCA SRIF-3} cluster {\sc MERLIN} in Cardiff.

\appendix\section{Disk definition}

In order objectively to identify disks and e.auxvaluate their masses and radii, we apply the following procedure. First, we isolate all the material with density $\rho >10^{-12}\,{\rm g}\,{\rm cm}^{-3}$. Second, we compute the moment of inertia tensor for this material, and thereby define a new local co-ordinate system using the principal axes of inertia, $x_{_{\rm IF}}$, $y_{_{\rm IF}}$, $z_{_{\rm IF}}$ (where the subscript "{\sc if}" stands for inertial frame).  The properties of the dense material are now analyzed relative to this new co-ordinate system. In particular, $z_{_{\rm IF}}$ is allocated to the largest principal moment of inertia, and therefore, if the material is in a disk, $z_{_{\rm IF}}$ is its rotation axis.

In order to ascertain whether there is a disk, we compute the logarithmic density profile along each of the axes of inertia, and smooth these profiles using a box-car averaging technique. For a disk, the profiles along $x_{_{\rm IF}}$ and $y_{_{\rm IF}}$ are very similar to one another, and the third, along $z_{_{\rm IF}}$ is significantly steeper and less extended.

Because disks are embedded in, and grow from.aux, filaments, we locate the edge of the disk at the first point along the $x_{_{\rm IF}}$- and $y_{_{\rm IF}}$-axes where the density is below $\rho_{_{\rm THRESH}}=10^{-14}\,{\rm g}\,{\rm cm}^{-3}$ {\it and} the second derivative of the logarithmic density profile is zero, $d^2\log_{_{10}}\rho/d(\log_{_{10}}r)^2=0$. Fig. A1 shows the profiles along the principal axes for the disk formed in simulation with $k_{_{\rm MIN}}\!=\!1/2$ and seed 400. This is the disk illustrated in the left column of the middle row of Fig. 2. All the disks in Fig. 2 have been identified in this way, and are viewed face-on down $z_{_{\rm IF}}$ in their local inertial frame.

The dotted vertical lines on Fig. A1 mark the radii at which the density first falls below $\rho_{_{\rm THRESH}}$, at $88\,{\rm  AU}$ on $x_{_{\rm IF}}$, and at $95\,{\rm AU}$ on $y_{_{\rm IF}}$, respectively. We identify the edge of the disk where the second derivative of the logarithmic density next falls to zero. This gives a mean radius of $R_{_{\rm DISK}}\!=\!94\,{\rm AU}$, in this case. The density typically drops very steeply inside this radius, and therefore the resulting estimate of the disk mass ($M_{_{\rm DISK}}\!=\!0.57\,{\rm M}_{_\odot}$ in this case) is robust.

\begin{figure}
\label{FIG_RHODISK}
\begin{centering}
    \psfig{figure=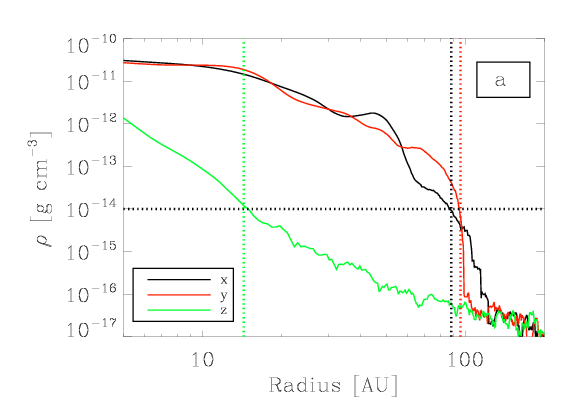, width=80mm}  
\caption{Density profiles along the principal axes in the local inertial frame for the disk formed in the simulation with $k_{_{\rm MIN}}\!=\!1/2$ and seed 400. The dotted vertical lines mark the radii at which the density along the $x_{_{\rm IF}}$- and $y_{_{\rm IF}}$-axes drops below the threshold density of $\rho_{_{\rm THRESH}}=10^{-14} \textrm{g cm}^{-3}$.}
\end{centering}
\end{figure}

\bibliographystyle{mn2e}
\bibliography{references}

\end{document}